\documentclass{jaa}

\def\aj{AJ}%
\def\araa{ARA\&A}%
\def\apj{ApJ}%
\def\apjl{ApJ~}%
\def\apss{Ap\&SS}%
\def\aap{A\&A~}%
\def\mnras{MNRAS~}%
\def\nar{New A Rev.}%
\def\sovast{Soviet~Ast.}%
\def\gca{Geochim.~Cosmochim.~Acta}%
\def\memsai{Mem.~Soc.~Astron.~Italiana}%
\usepackage{natbib}

\usepackage{graphicx}
\usepackage{caption}	
\usepackage{subcaption}	
\usepackage{float}
\usepackage{hyperref}
\hypersetup{colorlinks,linkcolor={black},citecolor={blue},urlcolor={red}}

\begin{document}\sloppy

\title{X-ray study of WR 48-6: A possible colliding wind binary.}


\author{Vishal Jadoliya\textsuperscript{1}, Jeewan C Pandey\textsuperscript{2} and Anandmayee Tej\textsuperscript{1,*}}
\affilOne{\textsuperscript{1}Indian Institute of Space Science and Technology, Thiruvananthapuram 695 547, Kerala, India\\}
\affilTwo{\textsuperscript{2}Aryabhatta Research Institute of Observational Sciences, Manora Peak, Nainital 263 129, India.\\}

\twocolumn[{

\maketitle

\corres{tej@iist.ac.in (AT)}

\msinfo{1 January 2015}{1 January 2015}

\begin{abstract}
This paper presents an investigation of the X-ray emission associated with the Wolf-Rayet star, WR 48-6, using observations from the {\it XMM Newton} and {\it Chandra} X-ray telescopes covering two epochs separated by eleven months. 
The X-ray spectrum of WR 48-6 is well explained by a two-temperature plasma model, with cool and hot plasma temperatures of $0.8_{-0.2}^{\,+0.1}$ and $2.86_{-0.66}^{\,+1.01}$ keV.
No significant X-ray variability is observed during these two epochs of observations. However, an increase in the local hydrogen column density accompanied by a decrease in the intrinsic X-ray flux between two epochs of observations is seen.
Additionally, the intrinsic X-ray luminosity is found to be more than $10^{33} \rm\,erg\,s^{-1}$ during both epochs of observations. Based on the analysis presented, WR 48-6 is a promising colliding wind binary candidate with a possible companion of spectral type O5--O6.

\end{abstract}


}]


\doinum{12.3456/s78910-011-012-3}
\artcitid{\#\#\#\#}
\volnum{000}
\year{0000}
\pgrange{1--}
\setcounter{page}{1}
\lp{1}

\section{Introduction}
Massive O-type stars evolve over short timescales ($\sim$ 10~Myr) into the Wolf-Rayet (WR) phase. The intense radiation fields of these WR stars expel and accelerate the outer layers, giving rise to powerful supersonic winds. These winds can reach terminal velocities ($v_{\infty}$) of 1000 -- 3000 \,\rm km\,s$^{-1}$ and result in very high mass loss rates ($\Dot{M}$), typically in the range of $10^{-7}-10^{-4}$ \,\rm M$_\odot$\,yr$^{-1}$ \citep{vanderHucht2001, Crowther2007}.
The line-driven wind mechanism, in which strong ultraviolet (UV) resonance lines absorb and scatter photons, produces a significant net outward force on the surrounding interstellar medium (ISM) from the photospheric radiation field.

X-ray emission from massive stars, especially in the WR phase, has been extensively studied in the last few decades, where the circumstellar environment is shown to be a key ingredient responsible for the observed X-ray emission.
Single WR stars are generally observed to be soft X-ray ($<$~1 keV) sources with X-ray to bolometric luminosity ratio ($L\rm _X/L\rm _{bol}$) of $\sim$ $10^{-7}$. It should, however, be noted that there are a few single WR stars which shows $L\rm _X/L\rm _{bol}$ $<$ 10$^{-7}$ \citep{Oskinova2016}.
Here, the emission originates from the high-velocity radiative shocks generated within the supersonic stellar wind through line-driven wind instabilities (\citealt{Lucy1980,Lucy1982,Owocki1988,Feldmeier1997}).
However, a few pointed observations of single WR stars also showed hard X-ray emission ($>$ 2 keV), contrary to expectations based solely on line-driven instability shocks, which typically do not predict hotter plasma \citep{Skinner2010,Skinner2012,Sokal2010}. However, single WC stars tend to be very faint or even dark in X-rays \citep[][]{ 2003A&A...402..755O,2006Ap&SS.304...97S}.

If the WR star is in a binary system (with another WR or OB companion), a hard X-ray ($>$~2 keV) component arises due to the strong hydrodynamic shocks in the wind collision region (WCR) that produce very hot ($> 10^6$~K) X-ray emitting plasma.
As a result, the X-ray luminosity in these colliding wind binary (CWB) systems is found to be higher than is expected from individual stars \citep[e.g.][]{Usov1992, Stevens1992}. Furthermore, the signature of periodic, phase-locked X-ray variability is observed in both long-period ($\sim$ few weeks to years) and short-period ($\sim$ few days) CWBs \citep{Luo1990,Usov1992,Antokhin2004,Pandey2014, Arora2019, Arora2020}. In long-period CWBs, the X-ray variability is attributed to the change in the binary separation, whereas, for short-period CWBs, it results from absorption due to variation in wind density along the line-of-sight towards the WCR.  

\begin{figure*}
     \centering
     \includegraphics[scale=0.65]{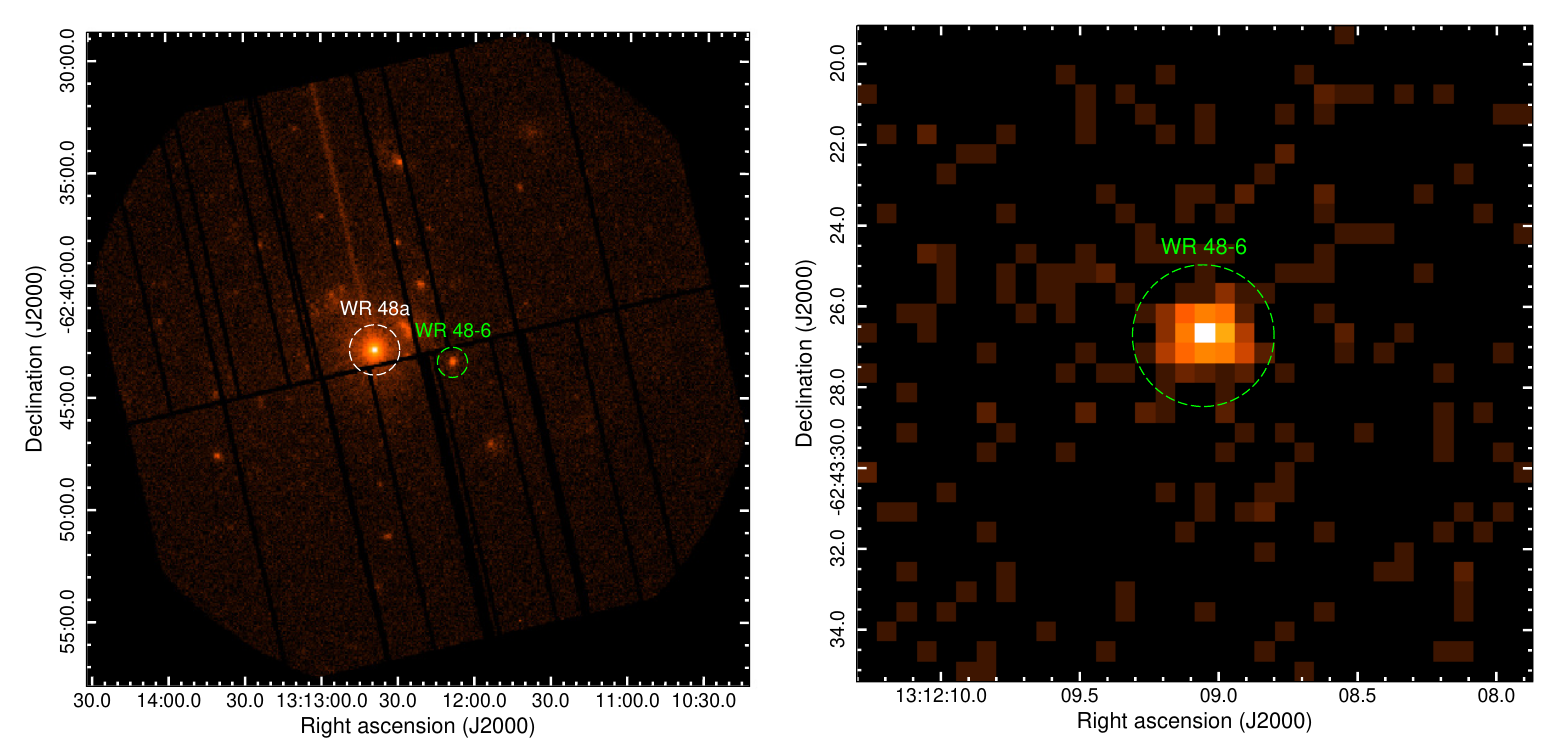}
         \caption{{\it Left:} {\it XMM-Newton} EPIC PN false colour image of the G305 field in the energy band 0.3--10.0 keV as observed by EPIC-PN of XMM-Newton. The locations of WR 48a and WR 48-6 are encircled. {\it Right:} {\it Chandra} ACIS-I false colour image of the G305 field in the energy band 0.5--7.5 keV.}
         \label{Fig:XMM-Chandra-Images}
\end{figure*}

\begin{table*}[htb]
\setlength{\tabcolsep}{1.2pt}
        \centering
        \tabularfont
	\caption{Log of observations of WR 48-6.}
	\label{tab:log}
	\begin{tabular}{lcccccccccr} 
		\hline
		Satellite & Observation & Observation Date  & Start time & Detector & Exposure & Bkg$^*$ Flare & Effective  & Source & Offset \\
		 &     ID     &  (YYYY:MM:DD) & UT(hh:mm:ss) & (Filter) & Time (s) & from to (ks) & Exposure (s) & counts & (arcmin) \\
		\hline
		XMM-Newton & 0510980101 & 2008 Jan 09  & 19:12:28   & MOS1 (M) & 71495 & 0-15.3 & 56128 & 602 & 3.541\\
	               &            &              &            & MOS2 (M) & 71508 & 0-15.3 & 56136 & 670     \\
		           &            &              &            & PN (M)   & 68965 & 0-17.4 & 51560 &1484 &      \\
	     Chandra   &    8922    & 2008 Dec 13  & 01:44:47   & ACIS-I   &119450 &        &       & 458 & 1.474\\
		\hline
	\end{tabular}
 \begin{flushleft}
\textbf{Note:}
$^*$ Background proton flare duration.
\end{flushleft}
\end{table*}

This paper presents a detailed analysis of the X-ray emission from a WR star, WR 48-6, using archival data obtained from {\it XMM-Newton} and {\it Chandra} telescopes. WR 48-6 (MDM3, MVM2011b, 2MASS J13120905-6243267, CXO J131209.0-624326) was discovered by \citet{Mauerhan2011}. WR 48-6  is classified as a WN9 star \citep{Rosslowe2015}. It is located in the vicinity of the giant star-forming H II region G305.4+0.1 in the G305 complex of the young and massive open star clusters Danks-1 and Danks-2 in the Scutum-Crux \citep{Mauerhan2011,Davies2012}. 
To adopt a distance to WR 48-6, we scrutinize the available values from the literature. The kinematic distance estimate to the G305 complex is $4.2\pm2.0$ kpc \citep{Davies2012}. Using the O8-B3 supergiant, D1-3, these authors determined the spectrophotometric distance to Danks 1 to be $3.48^{+0.91}_{-0.71}$ kpc. Further, a weighted mean of $4.16\pm0.6$ kpc is calculated for non-supergiants in Danks 1. This is also the distance considered in \citet{Rosslowe2015}. We have also checked the distance estimate from Gaia DR3 \citep{Gaia2022}. The measured parallax to WR 48-6 has a large uncertainty ($0.0552 \pm 0.1143$ mas), rendering an unreliable distance estimate. Hence, for the analysis presented in this study, we have adopted the distance of $4.16\pm0.6$ kpc.

The paper is structured as follows. Section \ref{Obs_reduction} briefly describes the data and the reduction methodology. Section \ref{timing_spectral_analysis} outlines the results of the timing and spectral analysis. The main results are discussed in Section \ref{discussion}, and Section \ref{conclusions} summarizes the same.

\section{Observations and Data reduction}\label{Obs_reduction}

Two epochs of observations of WR 48-6 are available and retrieved from the archives of {\it XMM-Newton} \citep{Jansen2001} and {\it Chandra} \citep{Weisskopf2000}. The log of observations is given in Table \ref{tab:log}. Observations were carried out with the European Photon Imaging Camera (EPIC) and Advanced CCD Imaging Spectrometer (ACIS-I) instruments of {\it XMM-Newton} and {\it Chandra}, respectively. Two of the XMM-Newton's X-ray telescopes are equipped with EPIC MOS CCD arrays \citep{Turner2001}, the third is equipped with a different CCD array called EPIC PN \citep{Struder2001}.
The EPIC instrument provides observations over a field of view (FOV) of 30~arcmin $\times$ 30~arcmin in the energy range of 0.15--12.0 keV with a spectral energy resolution (E\,/\,$\Delta$E) of 20--50 and angular resolution (i.e., FWHM) of 5--6~arcsec. The ACIS-I detector \citep{Garmire2003} in Chandra consists of four front illuminated CCDs, each having $1024\times1024$ pixels with a pixel size of $\approx$ 0.492~arcsec, which provides high spatial resolution ($\sim$ 1~arcsec) observations over a FOV of 16.9~arcmin $\times$ 16.9~arcmin with an energy resolution of 130 eV at 1.49 keV and 280 eV at 5.9 keV. Images from both datasets are shown in Figure \ref{Fig:XMM-Chandra-Images}.

The EPIC data were reduced with the standard {\it XMM-Newton} Science Analysis System (SAS version 20.0.0) using the latest Current Calibration Files (CCFs). These raw Observation Data Files (ODFs) of the EPIC-PN and EPIC-MOS detectors were processed through a pipeline by executing the respective SAS tasks {\sc epproc} and {\sc emproc}.  The data were further checked for high background proton flaring. The data from all three detectors were found to be affected by high background flare events, which were screened by generating a new Good Time Interval ({\sc gti}) event list.  The high background-affected times and background-subtracted source counts are also given in Table \ref{tab:log}.

\begin{figure*}[htb]
     \centering
     \begin{subfigure}{0.48\textwidth}
         \centering
         \includegraphics[width=\columnwidth]{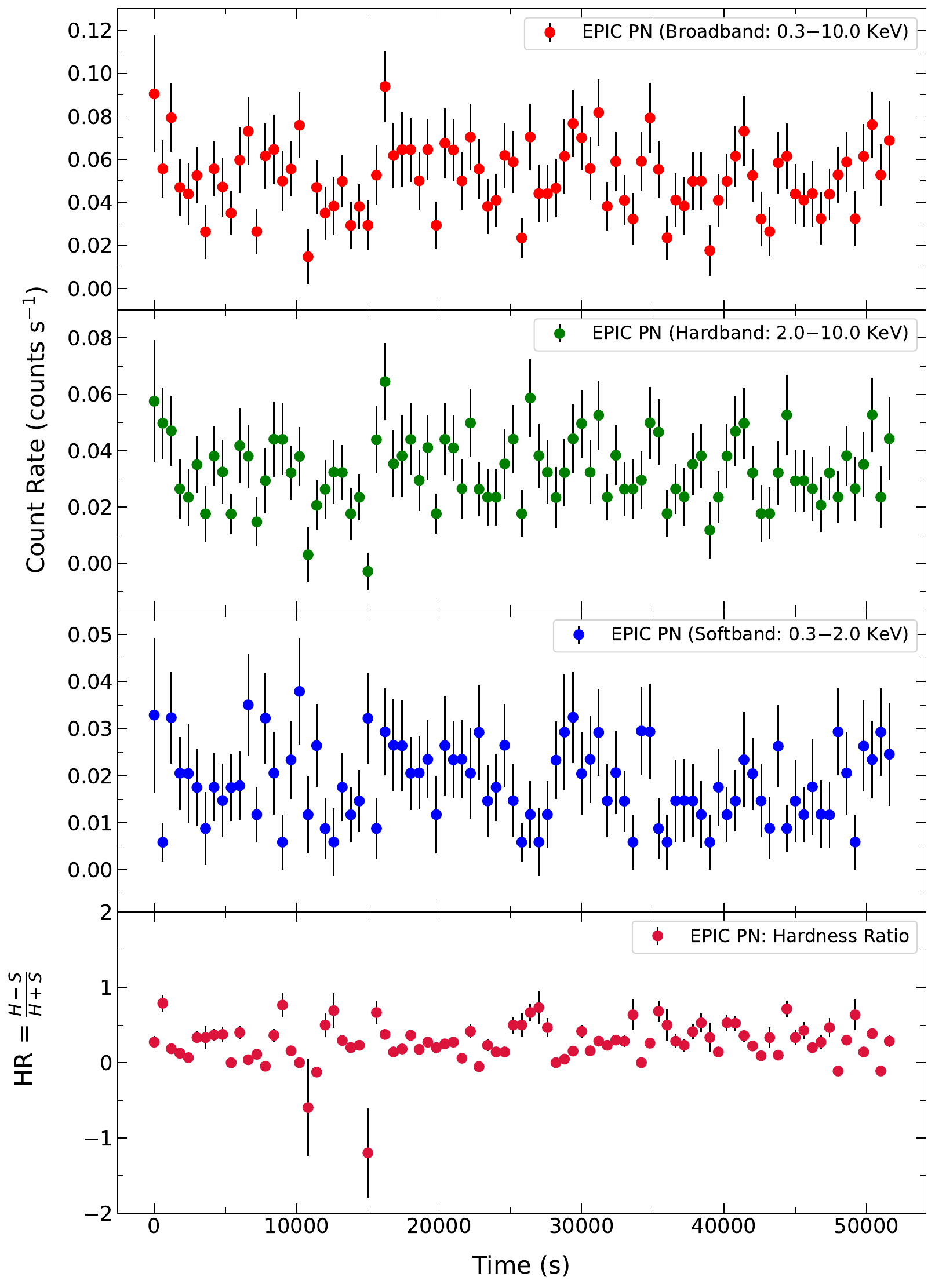}
         \caption{XMM-Newton}
         \label{XMM-Newton_LC}
     \end{subfigure}
     \begin{subfigure}{0.48\textwidth}
         \centering
         \includegraphics[width=\columnwidth]{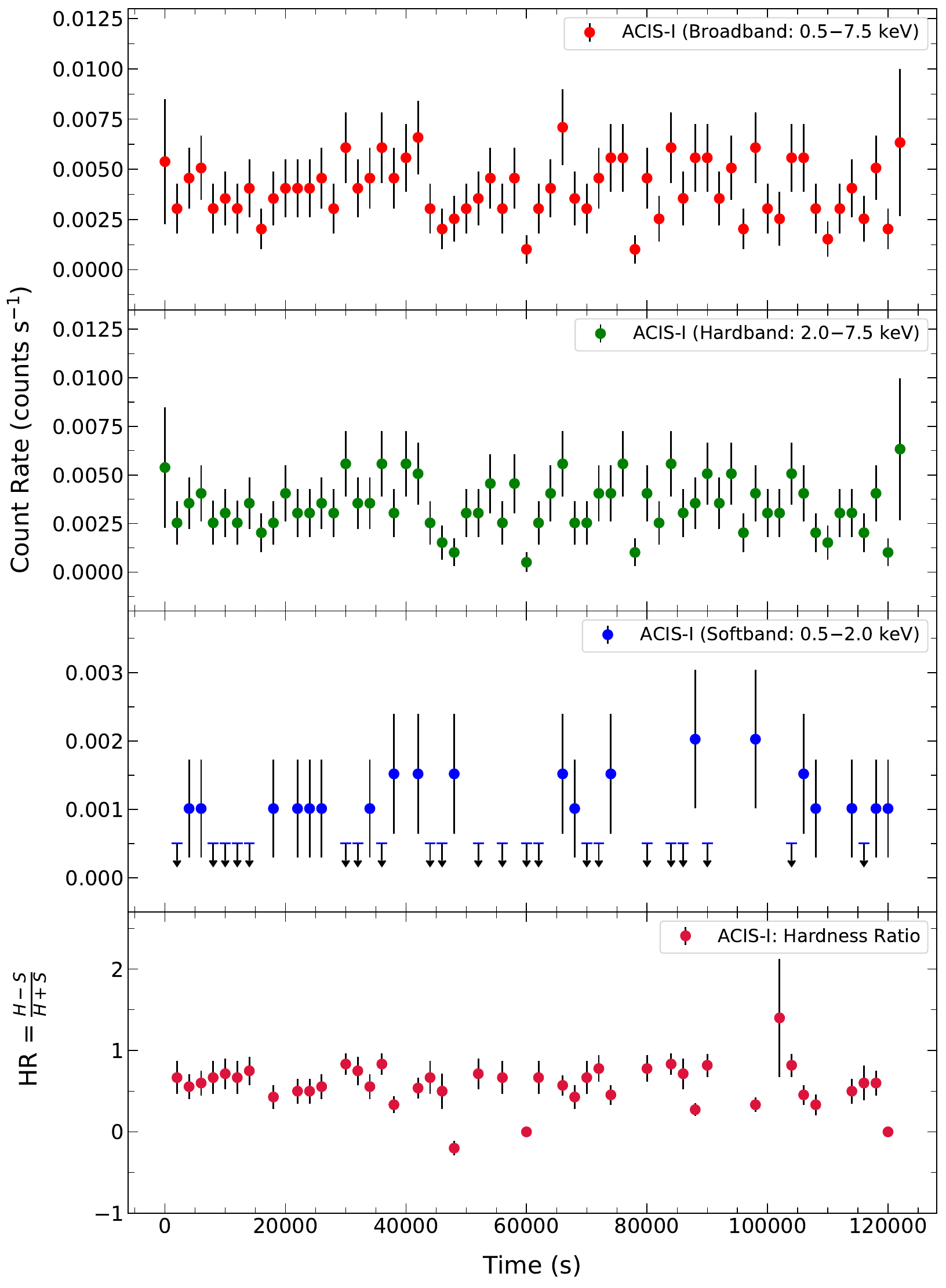}
         \caption{Chandra}
         \label{Chandra_LC}
     \end{subfigure}
\caption{The background subtracted X-ray light curves of WR 48-6 in total, hard, and soft energy bands along with the hardness ratio curve  observed with
 (a) EPIC-PN instrument of XMM-Newton and (b) ACIS-I instrument of Chandra. The bottom panels show the HR curves.}     
\end{figure*}

The pileup effect was also inspected using the task {\sc epatplot}, but neither MOS nor PN data was found to be affected by the pileup. The EPIC light curve and spectrum of WR 48-6 were extracted from a circular region of radius 20~arcsec, centred on the source, and the background regions were chosen from source-free regions around the source in the same CCD
of the detector.  
The task {\sc epiclcorr} was used to correct the light curves from the background and other effects.  Subsequently, the SAS task {\sc especget} was used to extract the source and background spectra, along with source-specific Ancillary Response Files (arf) and photon Redistribution Matrix Files (rmf). The EPIC spectra were grouped to have at least 20 counts in each spectral bin.
Further, we restricted the source's timing and spectral analysis within the energy range of 0.30--10.0 keV, because beyond this spectral range, the spectra were dominated by background and had insufficient count rate. 

{\it Chandra} data reduction was performed with the standard {\it Chandra} Interactive Analysis of Observations (CIAO) version 4.14 using the Calibration Database (CALDB) version 4.9.8, following the standard data reduction science threads. Level = 2 data files were generated by pipeline processing of raw, Level = 1 data files using the CIAO task {\sc chandra${\_}$repro}.
The source light curve and spectrum were extracted from a circular region of radius 1.5~arcsec, centred on WR 48-6, and the respective background correction files were extracted from a source-free circular region of the same radius near the source in the same CCD of the  ACIS-I.  
The task {\sc dmextract} was executed to generate the background subtracted X-ray light curves. Subsequently, the task {\sc specextract} was used to extract the spectra and the corresponding arf and rmf to calibrate the flux and energy in the spectral analysis of WR 48-6. Finally, the ACIS-I spectrum was grouped to have at least 10 counts per spectral bin. The timing and spectral analysis of WR 48-6 were restricted within the energy range of 0.5$-$ 7.5 keV as the counts were not significant beyond 7.5 keV.

\section{Timing and spectral analysis of WR 48-6}\label{timing_spectral_analysis}
\subsection{X-ray light curves analysis}
The {\it XMM-Newton} EPIC background-subtracted X-ray light curves were extracted in three energy bands: total (0.3--10.0 keV), hard (2.0--10.0 keV), and soft (0.3--2.0 keV). For {\it Chandra} ACIS-I, the extraction was in energy bands: total (0.5--7.5 keV), hard (2.0--7.5 keV) and soft (0.5--2.0 keV). The EPIC PN and ACIS-I light curves in the energy bands mentioned above were binned to 600~s and 2000~s time intervals, respectively, and are shown in Figures \ref{XMM-Newton_LC} and \ref{Chandra_LC}. 
     
To detect the presence of X-ray variability in WR 48-6, we have carried out the $\chi^2$ test on the X-ray light curves of each energy band obtained from EPIC and ACIS-I observations. The $\chi^2$ test examines the X-ray variability by calculating the significance of deviation for each data point of the light curve from and assumed constant count rate model, which is defined as follows:
\begin{equation}
    \chi^2=\sum_i^N\frac{(C_i-\bar{C})^2}{\sigma_i^2}
\end{equation}
where the X-ray photon count rate of $i^{th}$ observation is denoted by $C_i$, the mean count rate denoted by $\bar{C}$, which is used as the constant count rate model for the $\chi^2$ test, and $\sigma_i^2$ is the corresponding error in individual data points. 

\begin{table}
\setlength{\tabcolsep}{1.5pt}
    \centering
	\caption{ $\chi^2$-test results from X-ray light curves of WR 48-6.}
	\label{chi-test}
	\begin{tabular}{cccccc} 
		\hline
		 Detector & Energy & Mean count rate & $\chi^2$ & dof & P$_c$   \\
		          & band   &  (cts s$^{-1}$) &         &                &           \\
        \hline 
         EPIC PN & Total & 0.0516 $\pm$ 0.014  & 131.35 &  86 &0.0012  \\ 
                 & Hard  & 0.0330 $\pm$ 0.0111 & 131.87 &           &0.0011  \\
                 & Soft  & 0.0186 $\pm$ 0.0082 & 110.31 &           &0.0389  \\
         ACIS-I  & Total & 0.0040 $\pm$ 0.0014 &  91.70 &  61 &0.0067  \\ 
                 & Hard  & 0.0034 $\pm$ 0.0013 & 109.49 &           &0.0001  \\ 
                 & soft  & 0.0006 $\pm$ 0.0004 &    -   &    -      &  -     \\          
		\hline
	\end{tabular}
\end{table}

Next, the constant count rate probabilities (P$_c$) for individual light curves of {\it XMM-Newton} EPIC and {\it Chandra} ACIS-I were calculated for given values of $\chi^2$ and degree of freedom (dof). 
The results given in Table \ref{chi-test} indicate that these light curves were not constant during the observations. However, we did not see any periodic variability in the light curve of both epochs of observations.  Note that the $\chi^2$ test could not be performed for the MOS1, MOS2 light curves, and ACIS-I soft band light curve because of the very low count rate, and many of the data points are at the upper limit. 
In addition, we have estimated the hardness ratio (HR), which is defined as (hard-soft) / (hard + soft) for both the data from {\it XMM-Newton} and {\it Chandra}. The bottom panels of Figures \ref{XMM-Newton_LC} and \ref{Chandra_LC} show the HR curves. It appears that a significant amount of X-rays emanating from WR 48-6 exhibit energies above 2 keV.

\subsection{X-ray spectral analysis}\label{sec:xray_spectral}
\begin{figure*}
     \centering
     \begin{subfigure}{0.48\textwidth}
         \centering
     \includegraphics[width=\columnwidth]{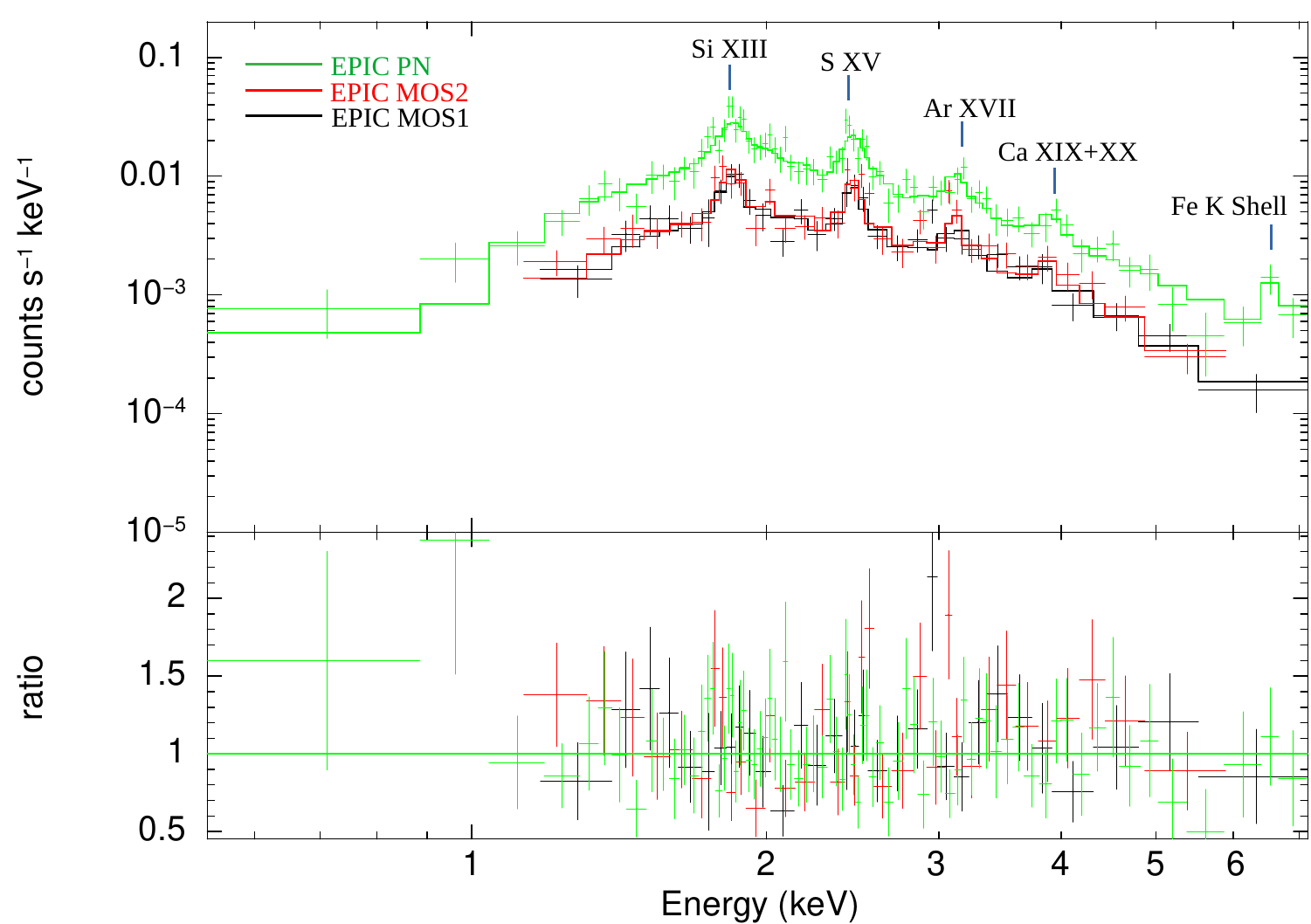}
     \caption{XMM-Newton}
         \label{Fig:XMM-Newton_sp}
     \end{subfigure}
     \begin{subfigure}{0.48\textwidth}
 \includegraphics[width=\columnwidth]{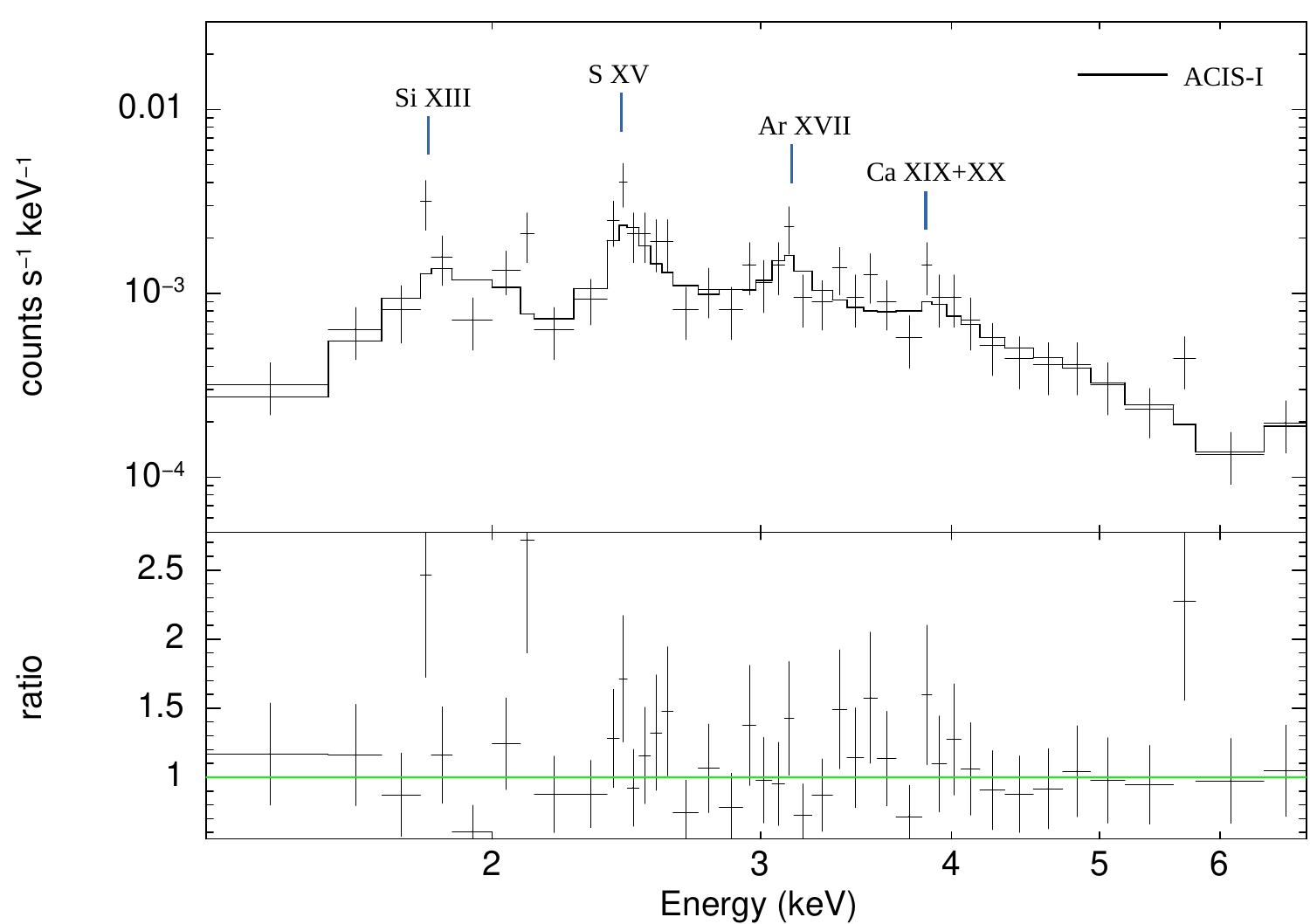}
 \caption{Chandra}
         \label{Fig:XMM-Newton_sp}
     \end{subfigure}
              \caption{X-ray spectra along with the best-fit, two-temperature thermal plasma model (2T-{\sc vapec}) for the observations from (a) {\it XMM-Newton} and (b) {\it Chandra}. The residual ratio is shown in the lower panel. The solid lines represent the best-fit model for the corresponding spectra.}
         \label{fig:sp}
\end{figure*}

The X-ray spectra of WR 48-6 observed by the \textit{XMM-Newton} EPIC and \textit{Chandra} ACIS-I detectors are shown in Figures \ref{fig:sp}.

The spectra observed in both epochs are affected by high absorption in the soft energy band ($<$~2.0 keV). Distinct emission lines of  Si XIII (1.853 keV), S XV (2.45 keV), Ar XVII (3.12 keV), Ca XIX+XX (3.9 keV) are identified in all spectra, where energies mentioned in the parentheses correspond to the mean energy for the He-like triplet/quartets.

\begin{table}
\centering
\caption{Best fit parameters from the X-ray spectral fitting. Errors are within a 90 per cent confidence interval} \label{tab:para}
\begin{tabular}{lcc}
\hline 
 Parameters                                                      & \multicolumn{2}{c}{Value}   \\ 
\cline{2-3}
                                                                 & EPIC                   & ACIS-I                   \\  
\hline 
 {$N\rm_H^{\rm local} \ (10^{22} \ \rm cm^{-2})$}                & $2.3_{-0.4}^{+0.4}$ & $3.8_{-0.8}^{+1.4}$   \\
 {$kT_1$ \ (keV)}                                                & $0.8_{-0.1}^{+0.1}$ & $1.0_{-0.4}^{+0.4}$   \\ 
 {$kT_2$ \ (keV)}                                                & $2.86_{-0.66}^{+1.01}$ & $ >  2.9$               \\ 
 {$EM_1 \ (10^{56} \ \rm cm^{-3})$}                              & $0.7_{-0.3}^{+0.4}$ & $0.2_{-0.1}^{+0.5}$   \\ 
 {$EM_2 \ (10^{56} \ \rm cm^{-3})$}                              & $0.14_{-0.06}^{+0.07}$ & $0.04_{-0.02}^{+0.04}$   \\ 
 Si                                                              & $3.3_{-1.0}^{+1.4}$ & Fixed*                   \\
 S                                                               & $6.8_{-2.1}^{+3.0}$ & Fixed*                   \\ 
 Ar                                                              & $13.5_{-5.7}^{+7.9}$ & Fixed*                   \\ 
 Ca                                                              & $9.9_{-6.5}^{+7.9}$ & Fixed*                   \\ 
 Fe                                                              & $2.4_{-1.3}^{+1.9}$ & Fixed*                   \\ 
 {$F\rm_S^{\rm obs} (10^{-13} \rm erg\ cm^{-2}\ s^{-1})$}        & $0.32\pm0.01$ & $0.036\pm0.003$                \\
 {$F\rm_H^{\rm obs} (10^{-13} \rm erg\ cm^{-2}\ s^{-1})$}        & $2.0\pm0.1$ & $0.65\pm0.05$                    \\
 {$F\rm_T^{\rm obs} (10^{-13} \rm erg\ cm^{-2}\ s^{-1})$}        & $2.3\pm0.1$ & $0.68\pm0.05$                    \\
 {$F\rm_S^{\rm ism} (10^{-13} \rm erg\ cm^{-2}\ s^{-1})$}        & $0.63\pm0.02$ & $0.063\pm0.005$                \\
 {$F\rm_H^{\rm ism} (10^{-13} \rm erg\ cm^{-2}\ s^{-1})$}        & $2.3\pm0.1$ & $0.73\pm0.06$                    \\
 {$F\rm_T^{\rm ism} (10^{-13} \rm erg\ cm^{-2}\ s^{-1})$}        & $2.9\pm0.1$ & $0.79\pm0.06$                    \\
 {$F\rm_S^{\rm int} (10^{-13} \rm erg\ cm^{-2}\ s^{-1})$}        & $30.2\pm1.0$ & $6.5\pm0.5$                     \\ 
 {$F\rm_H^{\rm int} (10^{-13} \rm erg\ cm^{-2}\ s^{-1})$}        & $3.4\pm0.1$ & $1.3\pm0.1$                      \\
 {$F\rm_T^{\rm int} (10^{-13} \rm erg\ cm^{-2}\ s^{-1})$}        & $33.6\pm1.2$& $7.9\pm0.6$                      \\ 
 {${\chi}^2/dof$}                                                & 122/134                & 35/38                 \\ 
             
\hline 
\end{tabular}
~\\
* Fixed as obtained from EPIC spectral fitting.\\
$F^{\rm obs}_{\rm S,H,T}$, $F^{\rm ism}_{\rm S,H,T}$, and $F^{\rm int}_{\rm S,H,T}$ are the observed, ISM corrected, and intrinsic flux, respectively in soft (S, 0.3--2.0 keV), hard (H, 2.0--10.0 keV) and total (T, 0.3--10.0 kev) energy bands.
\end{table} 

The EPIC PN spectra show the presence of the Fe K complex ($\sim$ 6.4--6.9~keV). Low count rates in the {\it Chandra} spectra could explain the non-detection of this line.
To derive the spectral parameters of WR 48-6, simultaneous spectral fitting was performed on the EPIC PN and MOS spectra in the 0.3--10.0~keV energy band. This was done by implementing models of the Astrophysical Plasma Emission Code \citep[apec;][]{Smith2001} in the X-ray spectral fitting software {\sc XSPEC} version 12.12.1 \citep{Arnaud1996}. The {\sc apec} model assumes that the X-ray emission is generated from optically thin plasma in collisional ionization equilibrium (CIE). X-ray emission from massive stars can be affected by the presence of interstellar material along the line of sight and the circumstellar medium around the star. Therefore, an X-ray absorption cross-section, modelled as the Tuebingen-Boulder ISM absorption \citep[tbabs;][]{Wilms2000}, was applied to correct the interstellar and local absorption effects along the line-of-sight by using the absorption components for the interstellar medium ($N\rm _H^{ISM}$) and the local hydrogen column density ($N\rm _H^{local}$). 
The observed X-ray spectra of WR 48-6 show emission lines over a wide range of energies, indicating that the CIE plasma might not be isothermal but could instead be a combination of cool and hot plasma.  
Considering this, we fitted a two-temperature plasma (2T-{\sc vapec}) model defined as {\sc tbabs*tbabs*(vapec+vapec)} and obtained the best-fit model for the observed X-ray spectra using a chi-square minimization algorithm. The interstellar medium hydrogen column density, $N\rm _H^{ISM}$, towards WR 48-6 was estimated from  \cite{HI4PI2016} and was found to be 9.1 $\times 10^{21} \rm cm^{-2}$. 
             
For modelling the EPIC spectra, $N\rm _H^{ISM}$ was fixed to the above value. Initially, we fixed the abundances to the canonical values for WN-type WR stars according to \citet{van1986}  with a comparison made against the solar photospheric values as per \cite{Anders1989}. Other parameters were left free to vary.  In doing so, the fit did not constrain the $N_H^{local}$ and temperatures of both components (kT$_1$ and kT$_2$) and was pegged to either lower or upper values.
The best fit was achieved by fixing the abundances of He, C, and N with those for WN9 type stars WR 108, as derived by \citet{Crowther1995}, with specific values relative to solar photospheric values: He/H = 6.82, C/H = 1.84, and N/H = 59.52. The abundance of O was set to the generic value for WN stars, O/H = 5.11 \citep{van1986}, while the abundances of Ne, Mg, Al, and Ni were fixed to match solar photospheric values.
 
The fitting algorithm then considers the parameters, $N\rm _H^{local}$, kT$_1$, kT$_2$ and the abundances of Si, S, Ar, Ca, and Fe, as free parameters. Modelling of the {\it Chandra} ACIS-I spectra, which have a low count rate, was done by fixing the abundances of Si, S, Ar, Ca, and Fe to the values estimated from the EPIC spectra. The best-fitted parameters obtained from the EPIC and ACIS-I spectra are given in Table \ref{tab:para}. 
The observed, ISM-corrected, and intrinsic X-ray fluxes of WR 48-6 were estimated in three energy bands: total, hard and soft. This was done using the {\sc cflux} model implemented in XSPEC. To calculate X-ray fluxes of ACIS-I data, the {\sc cflux} model was extended to the energy range of 0.3--10.0 keV by executing the task {\sc energies extend}. 
The emission measures EM$_1$ and EM$_2$, corresponding to cool and hot plasma components, were also estimated from the normalization parameters of {\sc vapec} model.
Within the derived uncertainties, the temperature values are consistent between the two epochs of observations. A similar trend is seen for EM$_1$, whereas a decrement by a factor of 3 is observed for EM$_2$ in the {\it Chandra} observation. Furthermore, the observed and ISM-corrected flux values in both soft and hard bands were found to be higher during the {\it XMM-Newton} observation, with the soft band flux an order of magnitude higher (see Figure \ref{fig:sp_var}). The elemental abundances estimated from the EPIC spectra were found to be non-solar. 

\section{Discussion}\label{discussion}
Investigation of the X-ray emission of WR 48-6, using {\it XMM-Newton} and {\it Chandra} observations, unravels a possible colliding wind binary scenario in this WN9 type star.  The evidence supporting this and discussion on the likely companion are presented in the following subsections.

\begin{figure}[htb]
     \centering
     \includegraphics[width=0.95\columnwidth]{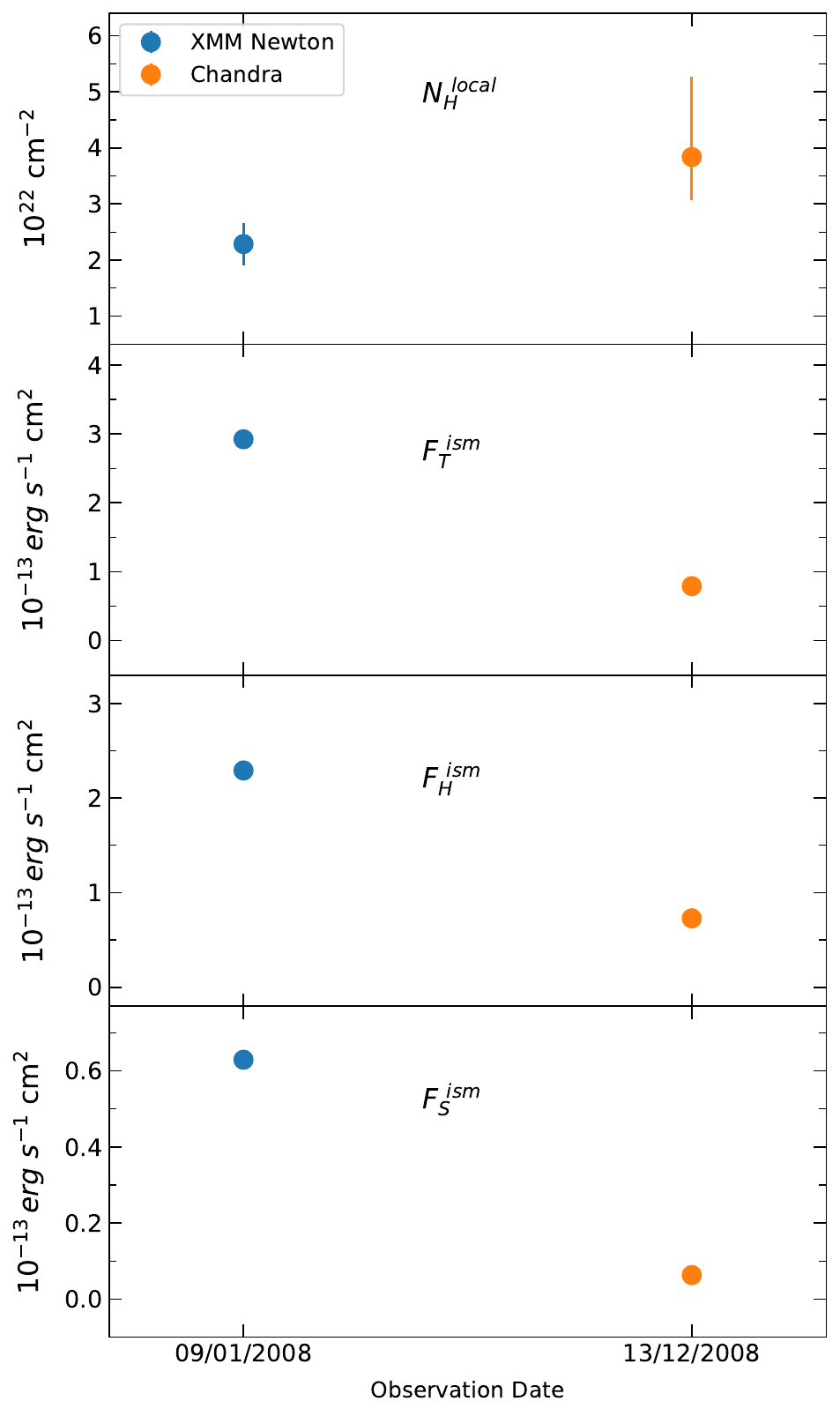}
         \caption{ Variation of $N\rm_H^{\rm local}$ and ISM corrected X-ray flux in the total, hard and soft energy bands in both epochs of observations of WR 48-6.} 
         \label{fig:sp_var}
\end{figure}

Based on {\it Chandra} data, \citet{Zhekov2017} has listed WR 48-6 as a binary system. In this section, we present a detailed investigation with the available archival data from {\it XMM-Newton} and {\it Chandra}.
The X-ray spectra obtained from the EPIC and ACIS-I observations in the energy bands 0.3--10.0~keV and 0.5--7.5~keV, respectively, provide a strong indication that the X-ray emission is generated from hot ($> 10^6\,K$), optically thin plasma. The X-ray spectra from EPIC-PN show the presence of an iron complex (Fe K-shell) emission line ($\sim$ 6.4--6.9~keV) along with emission lines originating from other ionized metal species like Si, S, Ar, and Ca. These emission lines are generally found in the X-ray spectra of WR stars \citep[e.g.][]{Pandey2014,Gosset2016,Arora2019,Arora2020,Arora2021}. 
The iron complex emission line, detected in the spectra, indicates a high-temperature ($\sim$10~MK) plasma. This line could possibly originate from the high-temperature WCRs in CWB systems. However, the detection of this line in WR 48-6 does not confirm the CWB scenario since it is also detected in the spectra of some single WR stars \citep{Skinner2010,Skinner2012}.

A simple two-temperature plasma model reasonably represents the X-ray spectra from EPIC and ACIS-I observations of WR 48-6. The average temperatures of the cooler and hotter plasma components are found to be 0.9~keV and  $\sim$ 2.9~keV, respectively. If we consider CWB systems, then the WCRs are temperature-stratified regions. Hence, in this two-temperature model, the high-temperature component can be interpreted to originate from the denser part of the WCR, compared to the low-temperature component, with the former being on the axis of symmetry and the latter away from it. However, in a few theoretical models \citep[e.g.][]{Cherepashchuk1976,Prilutskii1976,Usov1992}, the two temperature plasma components are interpreted as follows. 
The hotter plasma component, ranging from 2.0--4.0~keV, is believed to be produced in the WCR due to the collision of stellar winds arising from the binary system's primary (WR) and secondary components. On the other hand, the cooler component is found with a plasma temperature $\leq1.0$~keV that could be generated in the stellar winds of individual stellar components due to line-driven stellar wind instability shocks. 

The observed X-ray luminosity lends further support to the above interpretation. The X-ray luminosity of WR binary systems with a companion O-type star ranges from $\approx 10^{32}-10^{35} \rm erg\, s^{-1}$ \citep{Gagne2012}. The X-ray luminosity of WR 48-6, after correcting for absorption due to ISM ($L\rm_X^{ism}$), is calculated to be $6.1_{-1.8}^{+1.8} \times 10^{32}$ \rm erg s$^{-1}$  for EPIC and $1.6_{-0.5}^{+0.5} \times 10^{32}$\,\rm erg s$^{-1}$ for ACIS-I observation in the energy band of 0.3$-$10.0 keV. Additionally, the local and ISM-corrected X-ray luminosity ($L\rm_X^{int}$) is estimated to be $6.9_{-2.0}^{+2.0} \times 10^{33}$ \rm erg s$^{-1}$  for EPIC and $1.6_{-0.5}^{+0.5} \times 10^{33}$\,\rm erg s$^{-1}$ for ACIS-I observations. These estimated values of X-ray luminosity for WR 48-6 are not only consistent with the X-ray luminosity of WN+O CWBs \citep{Skinner2007,Zhekov2012}, but also higher than the X-ray luminosity observed in single WN-type WR stars \citep{Skinner2012,Oskinova2016,Naze2021}.
  
From spectral modelling, the ISM-corrected flux in the total band during the {\it XMM-Newton} observation is found to be more than 3 times that from the {\it Chandra} observation. A similar change in the flux was also found in other WN+O CWBs such as WR 139 \citep{Bhatt2010, Lomax2015}, WR 21a \cite{Gosset2016}, WR 25 \citep{Arora2019}, suggesting that WR 48-6 possibly belongs to the category of CWBs. Figure \ref{fig:sp_var} shows the variation of the $N\rm_H^{local}$ and the ISM-corrected fluxes. The increase in $N\rm_H^{local}$ accompanied by decreasing ISM-corrected flux further supports the CWB scenario.

Phase-locked variability in X-ray emission is another important feature that is observed in CWBs. As the companion star orbits around the primary WR star, periodic variation in the X-ray flux is observed due to the change in the local column density of material along the line of sight towards the observer. The high absorption of soft X-rays in the local shocked material decreases the observed X-ray flux, which increases the hardness of the X-ray emission and is also the cause of the periodic variations of the hardness in CWBs. 
In the case of WR 48-6, the light curves during both epochs of observations indicate the presence of variability. This observed variability could be attributed to either the presence of clumps in the wind or changes in the local absorption within the binary system, with the latter being particularly noticeable in the soft X-rays. However, the light curves show no evidence of cyclic variation within short time scales too, implying that any periodic patterns may extend beyond the duration of observation.

Furthermore, given that WR stars are mostly in binary systems, we attempt to constrain the properties of this system using the estimated values of $L\rm _X^{int}$ and  $N\rm_H^{local}$ with reasonable assumptions. The detailed discussion is presented in \ref{Binary-Comp}. The companion of WR 48-6 system is likely to be of spectral type O5--O6, though we are unable to constrain the luminosity class. Considering these spectral types for the companion, WR 48-6 appears to be a promising candidate for a close binary system  with a large shock opening angle and wind momentum ratio,  indicating radiative braking in the WCR. The proposed short orbital period also supports radiative cooling, matching the system's characteristics with other short-period CWBs. 
 
\vspace{-2em}
\section{Conclusions}\label{conclusions}
Detailed investigation of the X-ray emission from WR 48-6, a WN9 spectral type WR star, was carried out using the X-ray observations at two different epochs from the {\it XMM-Newton} and {\it Chandra} telescopes. Spectral analysis shows the presence of emission lines originating from ionized metal species like Si, S, Ar, and Ca, which are commonly observed in the spectra of WR stars. 

The local and ISM-corrected X-ray luminosity ($L\rm_X^{int}$) is estimated to be $6.9_{-2.0}^{+2.0} \times 10^{33}$ \rm erg s$^{-1}$ for EPIC and $1.6_{-0.5}^{+0.5} \times 10^{33}$\,\rm erg s$^{-1}$ for ACIS-I  observations, which are in good agreement with that of WR+O CWBs. Lending further support to the CWB scenario is the increase seen in $N\rm_H^{local}$ accompanied by a decline of the ISM-corrected flux from one epoch of observation to another. 
Evidence gathered from the timing and spectral study of the X-ray emission associated with WR 48-6 points towards a CWB system with a likely O5--O6 close companion. To gain a deeper understanding of the WR 48-6 system, a promising close CWB system, additional multiwavelength and multi-epoch observations are imperative.

\section*{Acknowledgements}
We thank the referee for valuable comments and suggestions that helped us improve the manuscript. The scientific outcomes presented in this research article have made use of data from the \textit{XMM-Newton} and \textit{Chandra}, including SAS, CIAO, and XSPEC software provided by the High Energy Astrophysics Science Archive Research Center (HEASARC), which is a service of the Astrophysics Science Division at NASA/GSFC.  We appreciate the reviewer of this paper for his/her careful reading of our paper and for providing valuable comments and suggestions.

\vspace{-1em}


\newpage
\appendix

\section{Possible Binary Companion }
\label{Binary-Comp}
Observationally, most WR stars in CWBs are seen to have O-type companions \citep{vanderHucht2001,DeBecker2013}. There are few WR + WR systems observed \citep[e.g.,][]{Callingham2020,Zhekov2014}. Therefore, from the observed statistics, it is more likely that an O-type star could be a companion of the WR star in WR 48-6 system. With the available data, we attempt to constrain the properties of this proposed binary system.
The pre-shock velocities of the stellar winds of the individual binary components of WR 48-6 are estimated by assuming the plasma temperatures  obtained from the X-ray spectral fitting (see Table \ref{tab:para}) as the post-shock temperature and using the following relation \citep{Luo1990}

\begin{equation}\label{eq:pre-shock}
 \centering
 kT_{\rm post-shock}=1.95\ \mu \ v_{\rm sh}^2  \,\rm keV
\end{equation}

where $\mu$ is the mean molecular weight of the outflowing gas and has the value of 1.16 \citep{Skinner2007} and 0.62 \citep{Cassinelli2008} for the WN and O-type star, respectively.  $ v_{\rm sh}$ is pre-shock stellar wind velocity in the units of 1000 $\rm km\,s^{-1}$. Table \ref{tab:sh} presents the results obtained for both stellar components. 
The pre-shock velocities corresponding to the cooler component of the plasma are found to be in the range of $\sim 500 - 800 \,\rm km\,s^{-1}$ and $\sim 600 - 900 \,\rm km\,s^{-1}$  for WN and O companion of WR48-6, respectively.
The average pre-shock velocities corresponding to the hot plasma component as obtained from both observations are found to be $\sim 1100 \,\rm km\,s^{-1}$ and $ \sim 1600 \,\rm km\,s^{-1}$ for WN and O-type companions of WR 48-6, respectively.

\begin{figure*}
     \centering
     \begin{subfigure}{0.49\textwidth}
         \centering
         \includegraphics[width=\textwidth]{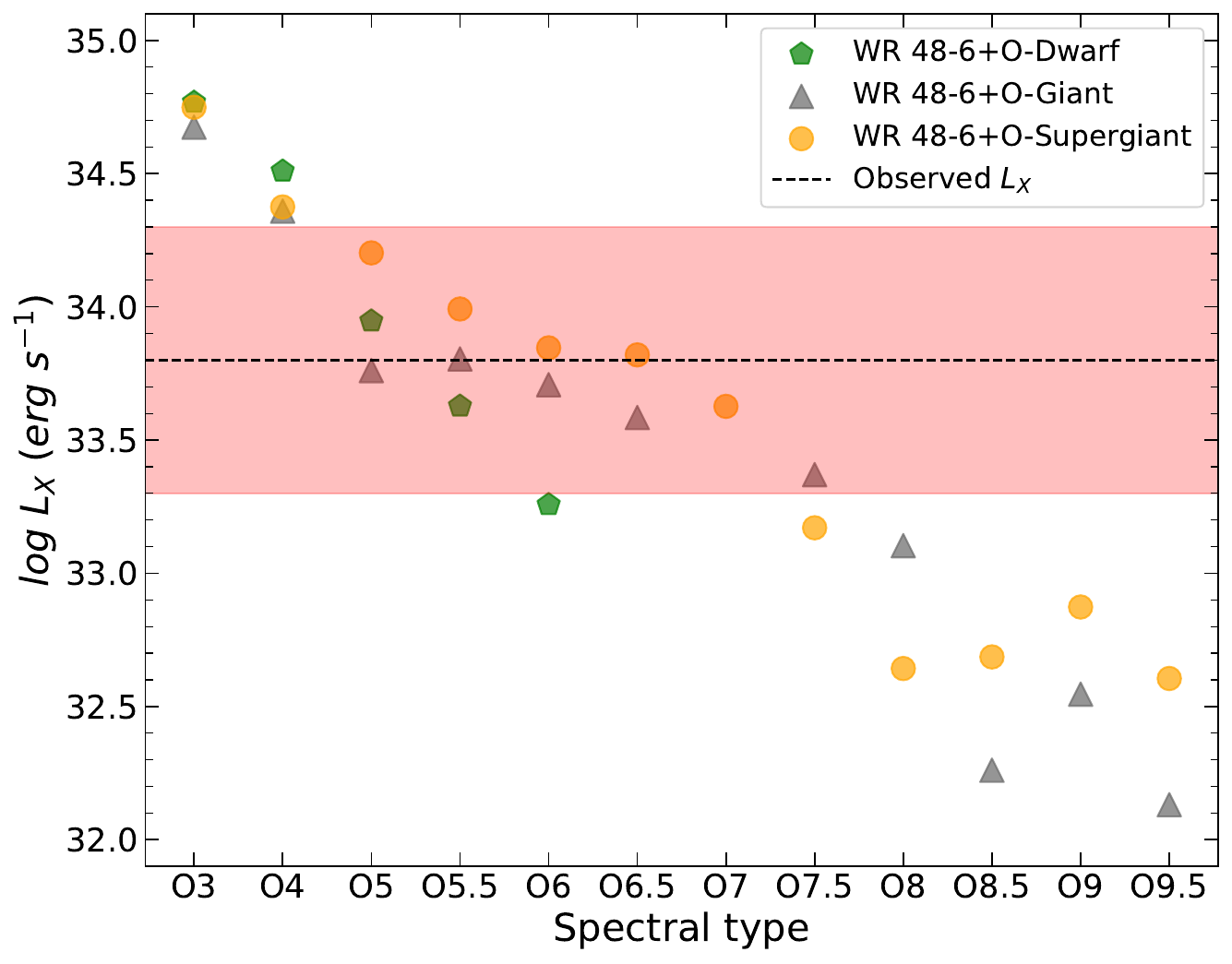}
         \caption{}
         \label{lxint}
     \end{subfigure}
     \begin{subfigure}{0.49\textwidth}
         \centering
         \includegraphics[width=\textwidth]{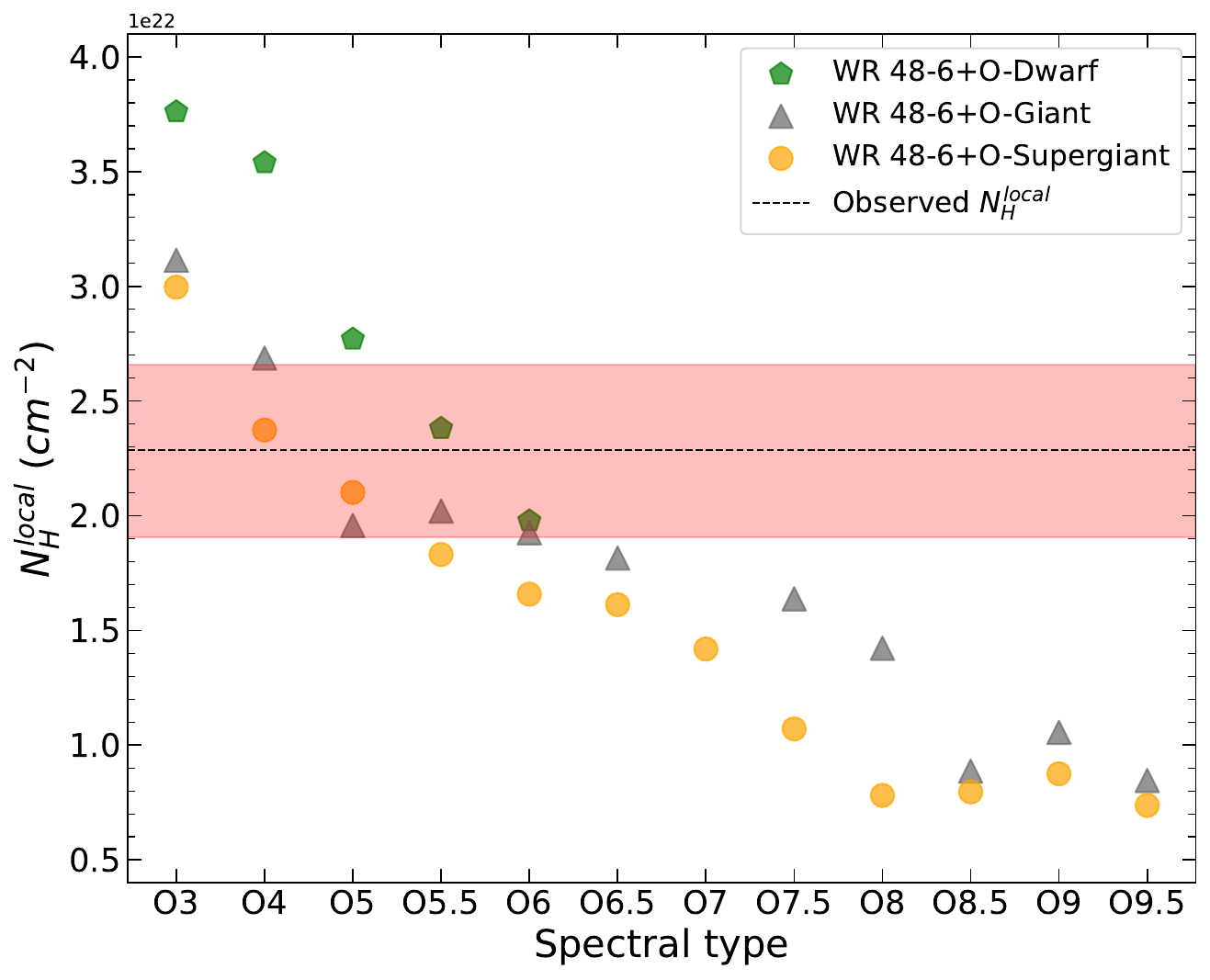}
         \caption{}
         \label{nhlocal}
     \end{subfigure}
\caption{(a) X-ray luminosity and (b) local hydrogen column density of WR 48-6 binary system considering WN9 as primary and O-type spectral type as companion star in the WR48-6 binary. The horizontal lines and shaded region around it represent the intrinsic X-ray luminosity and local hydrogen column density along their uncertainty within a confidence interval of 90 per cent obtained from the spectral fitting of XMM-Newton EPIC data.}
\label{fig:lx_nh}     
\end{figure*}

However, it is worth mentioning here that the WCR is confined by two oblique shocks that compress and heat the corresponding
stellar winds. Since the post-shock temperature is defined by the velocity component normal to the shock surface; thus, accurate estimation of the pre-shock velocity is difficult. Furthermore, as discussed earlier, WCRs are temperature (or density) stratified regions. Hence, the plasma temperatures derived from the spectral fitting can only be considered as a typical or average value.
Within the uncertainties imposed by the caveats mentioned above, these results qualitatively indicate that the stellar winds have reached their terminal speeds that could generate such high-temperature plasma in the WCR. Information about the mass-loss rate, the terminal velocity of stellar winds, and other vital parameters of WR 48-6 have not been estimated in the past. Therefore, we assumed typical values of these parameters as given by \citet{Crowther2007} for the corresponding WN spectral type of WR 48-6. However, the derived pre-shock velocity of WN companion of WR 48-6 is found to be higher than the typical value of the terminal velocity of the stellar wind mentioned in Table 2 of \citet{Crowther2007} for a WR star of WN9 spectral type. Therefore, we assumed the estimated pre-shock velocity ($\sim 1120 \,\rm km\,s^{-1}$) as the terminal velocity of the stellar winds of WR 48-6 ($v_{\infty}^{\rm WR}$)  for further analysis.

\begin{table}
    \centering
	\caption{Pre-shock velocities of WR 48-6 companions.} \label{tab:sh} 
\begin{tabular}{c c c c c}
\hline 
Observation & kT (keV) & \multicolumn{2}{c}{$v_{sh}$ $(\rm km \ s^{-1})$}  \\ 
\cline{3-4}
            &          &         WR       & O  \\
\hline

EPIC   & $0.8_{-0.1}^{+0.1}\, (\rm kT_1)$ & $\sim$590    &$\sim$ 810 \\ \\
       & $2.86_{-0.66}^{+1.01}\, (\rm kT_2)$ & $\sim$1120 & $\sim$1540  \\ \\
ACIS-I & $1.0_{-0.4}^{+0.3}\, (\rm kT_1)$ & $\sim$670  & $\sim$900 \\ \\
       & $> 2.9 \,\,\,\,\,\, (\rm kT_2)$  & $> 1130$           & $>1550$  \\ 

\hline 
\end{tabular}
 
\end{table}

In order to constrain the spectral type of the companion O-type star, the necessary parameters such as mass-loss rate ($ \dot{\rm M}_{\rm O}$), terminal velocity ($v_{\infty}^{\rm O}$), stellar radius ($\rm R_{\rm O}$), and the unitless parameter $\beta$ of O-type dwarf (V), giant (III), and supergiant (I) stars, are taken from \citet{Muijres2012}. It is noted that we have not considered those spectral types for which model parameters were not constrained \cite[see Table 1 of][]{Muijres2012}. The position of WCR from O-type companion ($\rm r_{\rm O}$) was estimated using the standard $\beta$-velocity law \citep{Castor1975,Abbott1982,Friend1986,Usov1992}, which is described as

\begin{equation}\label{beta}
   {\rm v (r_O)} = v_{\infty}^{\rm O}\left(1-\frac{\rm R_{\rm O}}{\rm r_{\rm O}}\right)^{\beta}
\end{equation}
Here, we assumed that  {$\rm v (r_O)$} has reached the pre-shock velocity of O-type star. Further, the binary separation (D) 
is estimated using the equation \citep[see][]{Usov1992}
 \begin{equation}
   {\rm D} = {\rm r_o} \left(1 +\frac{1}{\sqrt{\eta}}\right)   ~~{\rm and}~~~ \eta = \frac{\dot{\rm M}_{\rm O}\  v_{\infty}^{\rm O}}{\dot{\rm M}_{\rm WR} \ v_{\infty}^{\rm WR}}
 \end{equation}

Here, $\eta$ is wind momentum ratio and estimated for combinations of WN9 primary and possible O-type companion by assuming the mass-loss rate ($\dot{\rm M}_{\rm WR}$) for WN9 spectral type from \citet{Crowther2007}. The estimated values of D and $\rm r_{\rm WR}$ lie in the range of 50--275 R$_\odot$ and 31--224 R$_\odot$, respectively.

Finally, following the eq. (81) and (98) of \cite{Usov1992}, the intrinsic X-ray luminosity $L\rm _X^{\rm int}$ and local hydrogen column density $N\rm_H^{local}$ in WCR of a possible CWB system containing WR and an OV/OIII/OI star are estimated. 
Figures \ref{lxint} and \ref{nhlocal} show the plot of  $L\rm _X^{\rm int}$ and $N\rm_H^{local}$, respectively, as a function of the spectral type of the O-type companion of WR 48-6 binary system. The estimated values of $L\rm _X^{\rm int}$ and  $N\rm_H^{local}$ are in the range of $10^{32.1~ \text{--} ~34.8}$ erg s$^{-1}$ and $10^{21.9~ \text{--} ~22.6}$ cm$^{-2}$, respectively. The intrinsic X-ray luminosity obtained from modelling the {\it XMM-Newton} EPIC data and the observed  $N\rm_H^{local}$ value are also plotted in Figure  \ref{fig:lx_nh} along with the corresponding uncertainties within a confidence interval of 90 per cent. 
From the above plots, calculated theoretical values of $L\rm _X^{\rm int}$ for O5--O7.5III, O6--O7I and O5V fall within the quoted uncertainties of the values derived from EPIC spectra (see Figure \ref{lxint}). If we compare the $N\rm_H^{local}$ values, then the values for O4--O6III, O4--O5I, and O5.5--O6V fall within the quoted uncertainties of the observed value. Based on the above, the spectral type of the companion of WR 48-6 can be O5--O6, though constraining the luminosity class is not possible.

The orbital period ($P_{\rm o}$) is an important parameter for a binary system and can be estimated by using Kepler's third law, which is expressed as
\begin{equation}
    P_{\rm o}^2=\frac{4\pi^2D^3}{G(M_O+M_{WR})}
\end{equation}
where $M_{\rm WR}$ and $M_{\rm O}$ are the masses of the WR star and its O-type companion, and D is the orbital separation between the binary components. Taking luminosity and mass-loss rate values for WN9 spectral type from  \citet{Crowther2007}, the mass of WR 48-6 ($M_{\rm WR}$) is estimated to be 17.9 $M_{\odot}$ using the mass-luminosity relation of \citet{Maeder1987}. From the constraints on the spectral type of the companion and assuming circular orbit, we deduce the orbital period to be in the range of 9--20 days, suggesting WR 48-6 to be a close binary system.

Cooling of the shock-heated plasma also plays an important role in the wind-wind collision of CWB systems. The gas present in the WCR either cools by emitting radiation, known as radiative cooling or due to the WCR's expansion, known as adiabatic cooling. A cooling parameter, $\chi$, is introduced to find the nature of the cooling in CWB systems, which is defined as the ratio of the time taken by post-shock gas to cool ($t_{\rm cool}$) to the time taken by the gas to escape ($t_{\rm escape}$) the WCR \citep{Stevens1992}
\begin{equation}
    \chi=\frac{t_{\rm cool}}{t_{\rm escape}}=\frac{v^4 \ d}{\dot{M}}
\end{equation}
where the pre-shock wind velocity ($v$), distance ($d$) of the WCR from the companion star, and mass-loss rate ($\dot{M}$) are in units of 1000 $\rm km \ s^{-1}$, $10^{7}$ \rm km, and $10^{-7} \ \rm M_{\odot}\,yr^{-1}$, respectively. In CWB systems, radiative cooling takes place when $\chi<<1$ and the gas cools adiabatically when $\chi>>1$. From the analysis, it is observed that the stellar winds of the WR 48-6 are consistent with radiative cooling.
We have also estimated the shock opening angle ($\theta$) to explore the nature of WCR in the WR 48-6 system using the expression \citep{Eichler1993}

\begin{equation}
    \theta = 2.1 \left(1-\frac{\eta^{2/5}}{4}\right){\eta^{1/3}}     \,\,\,\, \rm for \,\,\,10^{-4} \leq \eta \leq 1.0
\end{equation}

$\theta$ of spectral type O5--O6 for respective luminosity classes (V/III/I) is estimated to lie in the range of $44^\circ-58^\circ$, $55^\circ-69^\circ$, $68^\circ-76^\circ$ with large wind momentum ratios ($\eta$) in the range, 0.06--0.17, 0.14--0.32 and 0.30--0.45 for dwarf, giant and supergiant companion, respectively. These large values of $\theta$ and $\eta$ indicate radiative breaking in the WCR \citep{Gayley1997,Stevens1994} and can be observed for early O-type companion star, which agrees well with the spectral range of the O-type companion inferred.
The estimated short orbital period supports radiative cooling in close binary systems ($\chi \propto P_{\rm o}^{2/3}$) \citep{Stevens1992}. The signature of radiative braking due to a large shock opening and high wind momentum ratio found in the WR 48-6 system is similar to other short-period CWBs like WR 139 \citep{Marchenko1994,Bhatt2010,2015A&A...573A..43L} and WR 121a \citep{Arora2020}.

\vspace{-1.5em}
\end{document}